\documentclass[a4paper,11pt]{article}
\pdfoutput=1 

\usepackage{jinstpub} 

\title{\boldmath The Cylindrical GEM Inner Tracker of the BESIII experiment: prototype test beam results}

\author[a,f]{L. Lavezzi\note{ Corresponding author.}}

\author[f]{M. Alexeev}
\author[f,l]{A. Amoroso}

\author[a,c]{R. Baldini Ferroli}
\author[c]{M. Bertani}
\author[b]{D. Bettoni} 
\author[f,l]{F. Bianchi}

\author[c]{A. Calcaterra}
\author[b]{N. Canale}
\author[c,e]{M. Capodiferro}
\author[b]{V. Carassiti}
\author[c]{S. Cerioni}
\author[a,f,h]{JY. Chai}
\author[b]{S. Chiozzi}
\author[b]{G. Cibinetto}
\author[f,h]{F. Cossio}
\author[b]{A. Cotta Ramusino}

\author[f,l]{F. De Mori}
\author[f,l]{M. Destefanis}
\author[c]{J. Dong}

\author[b]{F. Evangelisti}

\author[b,i]{R. Farinelli}
\author[f]{L. Fava}
\author[c]{G. Felici}
\author[b]{E. Fioravanti}

\author[b,i]{I. Garzia}
\author[c]{M. Gatta}
\author[f,l]{M. Greco}

\author[a,f,h]{CY. Leng}
\author[a,f]{H. Li}

\author[f,l]{M. Maggiora}
\author[b]{R. Malaguti}
\author[f,l]{S. Marcello}
\author[b]{M. Melchiorri}
\author[b,i]{G. Mezzadri}
\author[f]{M. Mignone}
\author[c]{G. Morello}

\author[d,k]{S. Pacetti}
\author[c]{P. Patteri}
\author[f,l]{J. Pellegrino}
\author[c,e]{A. Pelosi}

\author[f]{A. Rivetti}
\author[f]{M. D. Rolo}

\author[b,i]{M. Savri\'e}
\author[b,i]{M. Scodeggio}
\author[c]{E. Soldani}
\author[f,l]{S. Sosio}
\author[f,l]{S. Spataro}

\author[c,g]{E. Tskhadadze}

\author[i]{S. Verma}

\author[f]{R. Wheadon}

\author[f]{L. Yan}

\affiliation[a]{Institute of High Energy Physics, Chinese Academy of Sciences, 19B YuquanLu, Beijing, 100049, China}
\affiliation[b]{INFN, Sezione di Ferrara, via G. Saragat 1, 44122 Ferrara, Italy}
\affiliation[c]{INFN, Laboratori Nazionali di Frascati, via E. Fermi 40, 00044 Frascati (Roma), Italy}
\affiliation[d]{INFN, Sezione di Perugia, via A. Pascoli 14, 06123 Perugia, Italy}
\affiliation[e]{INFN, Sezione di Roma, c/o Universit\`a La Sapienza, p.le A. Moro 2, 00185 Roma, Italy}
\affiliation[f]{INFN, Sezione di Torino, via P. Giuria 1, 10125 Torino, Italy}
\affiliation[g]{Joint Institute for Nuclear Research (JINR), Joliot-Curie 6, Dubna, Moscow region, 141980, Russia}
\affiliation[h]{Politecnico di Torino, Dipartimento di Elettronica e Telecomunicazioni, Corso Duca degli Abruzzi 24, 10129 Torino, Italy}
\affiliation[i]{Universit\`a di Ferrara, Dipartimento di Fisica, via G. Saragat 1, 44122 Ferrara, Italy}
\affiliation[k]{Universit\`a di Perugia, Dipartimento di Fisica e Geologia, via A. Pascoli 14, 06123 Perugia, Italy}
\affiliation[l]{Universit\`a di Torino, Dipartimento di Fisica, via P. Giuria 1, 10125 Torino, Italy}

\emailAdd{lia.lavezzi@to.infn.it}
\emailAdd{lavezzi@ihep.ac.cn}

\abstract{A cylindrical GEM detector is under development, to serve as an upgraded inner tracker at the BESIII spectrometer. It will consist of three layers of cylindrically-shaped triple GEMs surrounding the interaction point. The experiment is taking data at the e$^+$e$^-$ collider BEPCII in Beijing (China) and the GEM tracker will be installed in $2018$. \\
Tests on the performances of triple GEMs in strong magnetic field have been run by means of the muon beam available in the H4 line of SPS (CERN) with both planar chambers and the first cylindrical prototype. Efficiencies and resolutions have been evaluated using different gains, gas mixtures, with and without magnetic field. The obtained efficiency is $97 - 98\%$ on single coordinate view, in many operational arrangements. The spatial resolution for planar GEMs has been evaluated with two different algorithms for the position determination: the charge centroid and the micro time projection chamber ($\mu$-TPC) methods. The two modes are complementary and are able to cope with the asymmetry of the electron avalanche when running in magnetic field, and with non-orthogonal incident tracks. With the charge centroid, a resolution lower than $100$ $\mu$m has been reached without magnetic field and lower than $200$ $\mu$m with a magnetic field up to $1$ T. The $\mu$-TPC mode showed to be able to improve those results. \\
In the first beam test with the cylindrical prototype, the detector had a very good stability under different voltage configurations and particle intensities. The resolution evaluation is in progress.}


\collaboration{CGEM-IT group}

\proceeding{Instrumentation for Colliding Beam Physics (INSTR-17) \\
27 February 2017 - 3 March 2017 \\
Budker Institute of Nuclear Physics}

\begin{document}
\maketitle
\flushbottom

\section{Introduction}
\begin{figure}[htbp]
\centering 
\raisebox{1.5cm}{\includegraphics[width=.4\textwidth]{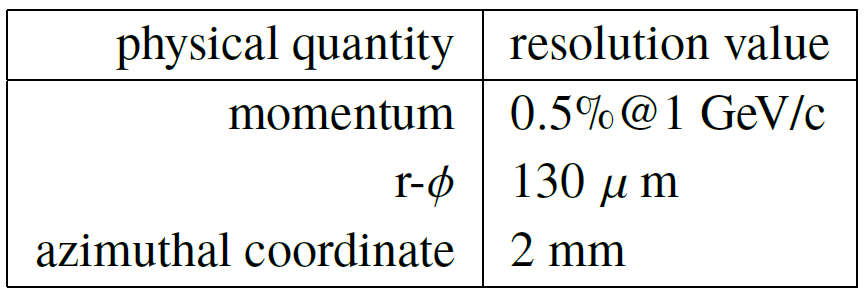}}
\qquad
\includegraphics[width=.5\textwidth]{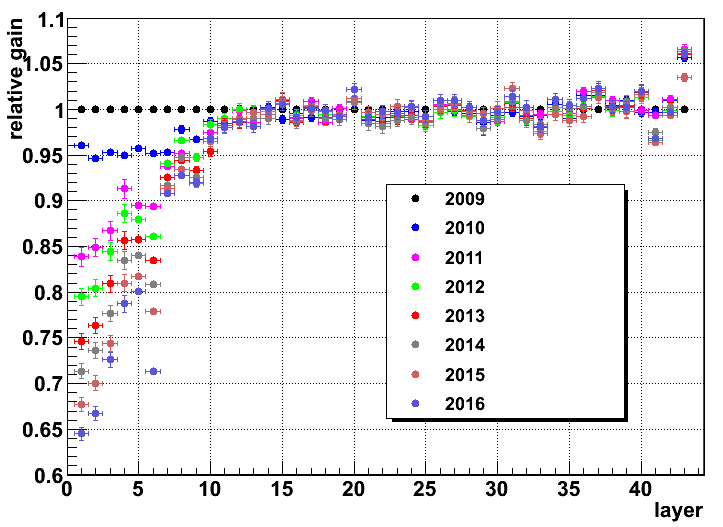}
\caption{MDC tracker performances ({\it left}); gain relative loss per year of the MDC tracker ({\it right}) \cite{aging}.} \label{fig:mdc} 
\end{figure}

The CGEM-IT will be the Inner Tracker of the BESIII experiment \cite{besiii} (Beijing, China) by $2018$. \\
The {\bf BE}ijing {\bf S}pectrometer {\bf III} (BESIII) has been taking data since $2009$ at the e$^+$e$^-$ {\bf B}eijing {\bf E}lectron {\bf P}ositron {\bf C}ollider {\bf II} (BEPCII, \cite{bepcii}). It is a charm and $\tau$ factory, with the energy in the center of mass in the range of $2 - 4.6$ GeV. In $2016$, the design luminosity of $1 \cdot 10^{33}$ cm$^{-2}$ s$^{-1}$ was reached. \\
The current tracking system, the {\bf M}ain {\bf D}rift {\bf C}hamber (MDC), is composed by an Inner and an Outer Tracker, sharing the same helium based gas mixture. In fig.~\ref{fig:mdc} ({\it left}) the performances of the MDC are summarized.
Due to the increasing luminosity, the MDC has been exposed to a radiation dose sufficient to produce an aging problem (fig.~\ref{fig:mdc} ({\it right})). The relative gain loss of the innermost layers is around $4\%$ per year. To cope with the potential MDC Inner Tracker breakdown, its substitution with a {\bf C}ylindrical {\bf G}as {\bf E}lectron {\bf M}ultiplier {\bf I}nner {\bf T}racker (CGEM-IT) has been proposed and accepted by the collaboration.
The CGEM-IT consists of three layers of cylindrical triple-GEMs. It can provide the 3D position of the charged particle at each plane, since the anode has strips both parallel to the cylinder axis and tilted of a large stereo angle ($> 30^\circ$). The CGEM-IT foresees an improved (w.r.t MDC) $z$ coordinate resolution (better than $1$mm), granting the same r-$\phi$ spatial and momentum resolutions. The outcome will be a more precise reconstruction of secondary vertices. The new tracker will guarantee the data taking up to the desired deadline of $2022$. \\
Other key points of the CGEM-IT are the very low material budget (X$_0 < 1.5\%$), the capability to sustain the high particle rate ($\sim$$10^4$ Hz/cm$^2$) and the increased angular coverage ($93\%$).
\begin{figure}[htbp]
\centering 
\includegraphics[width=.75\textwidth]{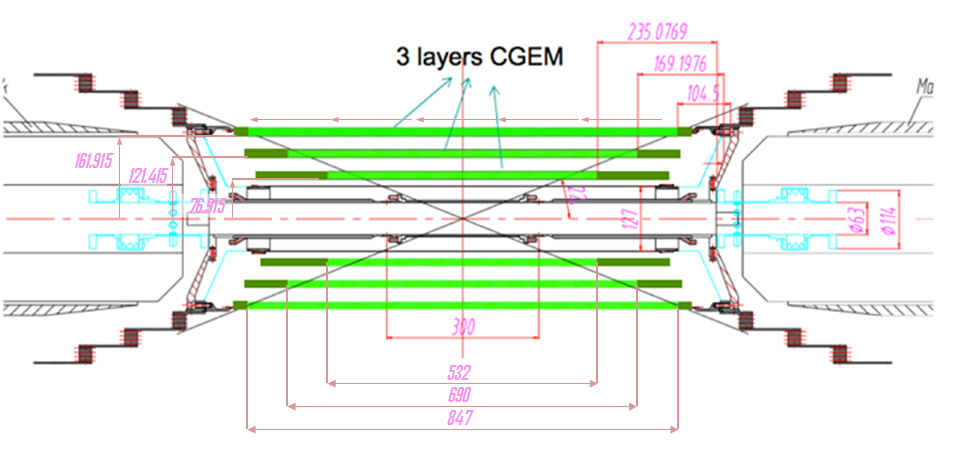}
\caption{Scheme of the BESIII CGEM-IT (measures are in mm).} \label{fig:cgem-it}
\end{figure}
\section{The GEM detector}
As already stated, the CGEM-IT is a cylindrical-triple-GEM detector. \\
The {\bf G}as {\bf E}lectron {\bf M}ultiplier (GEM) is a distributed flat electron amplifier in gas invented in 1997 by F. Sauli \cite{sauli}. A detector based on GEM operates in general as a standard gas tracker, i.e. a charged particle ionizes the gas, the produced electrons drift toward the anode, undergo avalanche multiplication and produce the signal which is readout. The peculiarity stands in the multiplication stage. The GEM foil is a metal coated polymer ($50$$\mu$m kapton $+$ $3$$\mu$m copper), pierced with holes of $\sim$$50$$\mu$m diameter. A voltage of some hundreds of Volts is applied between the two copper layers and the whole foil is immersed in an electric field (to produce the electron drifting). The voltage applied to the GEM sides is enough to create an intense electric field (some tens of kV/cm) in the very small holes and generate the avalanche multiplication of the electrons drifting through them. Gains up to $10^4$ are reachable with moderate voltages. \\
By three GEM foils between anode and cathode, instead of one, the same (or higher) gain values can be obtained, operating the GEMs at lower voltages, thus decreasing the discharge rate \cite{triple}.
\subsection{The CGEM: history and new features in BESIII}
The very first cylindrical triple GEM belongs to the KLOE-2 experiment (Frascati) \cite{kloe2}. BESIII borrows the construction procedure from it, with some improvements. The GEM and electrode foils are produced in plane and then shaped on cylindrical molds. The assembly is performed inside the Vertical Inserting Machine, built specifically for KLOE-2 and conveniently modified for BESIII. \\
BESIII will possess the first CGEM working with analog readout inside a magnetic field, providing the simultaneous measurement of the deposited charge and the time of arrival of the signal. \\ 
More original features are present. The adoption of Rohacell 31 instead of Honeycomb as support of the anode/cathode and the placement of the permaglass rings only outside the active area minimize the material budget. A dedicated ASIC is being developed: the {\bf T}orino {\bf I}ntegrated {\bf G}EM {\bf E}lectronics for {\bf R}eadout (TIGER) \cite{roloINSTR17}. A jagged strip anode is used, to lower the inter-strip capacitance \cite{isaTIPP14}.
\section{The test beam measurements}
Both planar chambers and the first cylindrical prototype have been tested on the H4 line of the SPS (North Area, CERN) \cite{H4}.
\begin{figure}[htbp]
\centering 
\includegraphics[width=.35\textwidth]{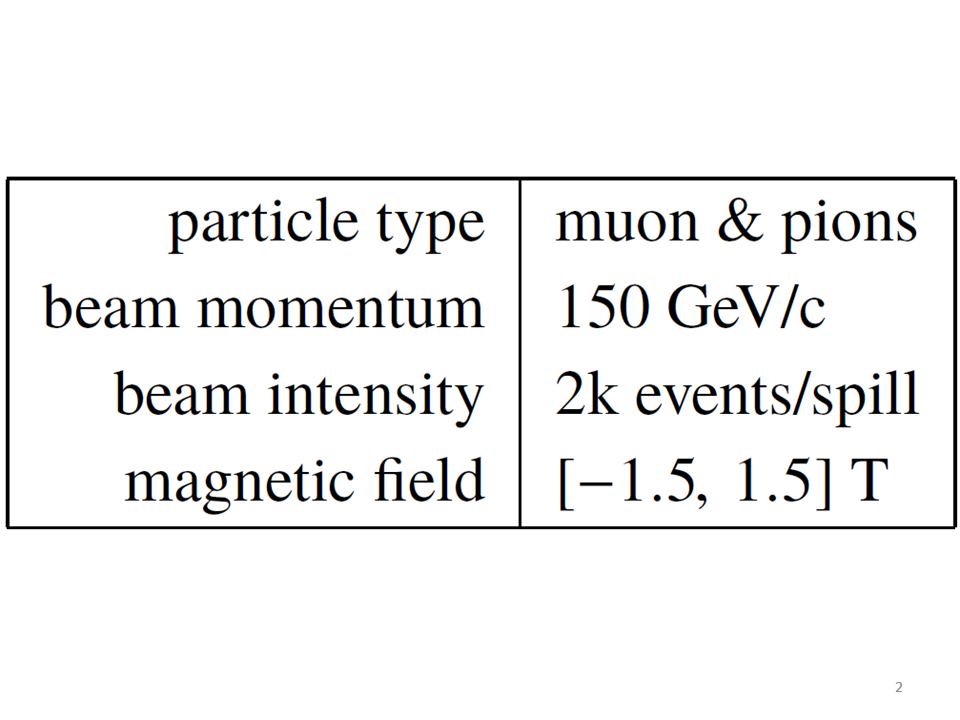}
\qquad
\includegraphics[width=.2\textwidth]{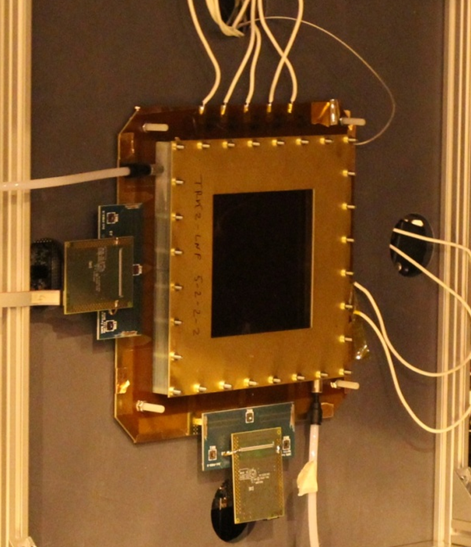} 
\qquad
\includegraphics[width=.3\textwidth]{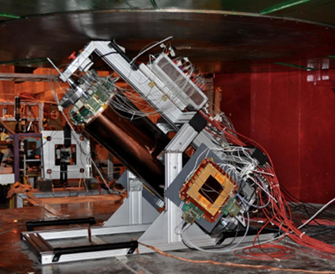}
\caption{H4 environment ({\it left}); a planar GEM ({\it center}); first test beam setup of the CGEM ({\it right}).} \label{fig:testbeam}
\end{figure}
The planar chambers are $10 \times 10$ cm$^2$ triple GEMs, with $x/y$ view readout and $3$mm or $5$mm drift gap. Ar:CO$_2$ ($70/30$) and Ar:i-C$_4$H$_{10}$ ($90/10$) gas mixtures were used. The cylinder is the first prototype of the Layer 2, with $3$mm drift gap, filled with Ar:i-C$_4$H$_{10}$ ($90/10$). \\
The main goal of the test beam measurements was to evaluate efficiencies and resolutions, to study the different high voltage and field settings in order to decide the working point and the gas mixture.
\subsection{Track reconstruction methods}
Two modes for track reconstruction are available: the charge centroid and the micro-TPC.
\paragraph{The charge centroid }
In the case of a Gaussian charge distribution on the anode, the weighted average of the firing strip positions is enough to evaluate the $x$ position of the charged track\footnote{on the anode.}:
\begin{equation} \label{eq:cc}
< x > = \frac{\sum_i x_i q_i}{\sum_i q_i}
\end{equation}
\paragraph{The micro-TPC ($\mu$-TPC)}
When the charge distribution is not Gaussian, the charge centroid method fails and another solution is needed: the $\mu$-TPC mode \cite{mu-TPC}. In this case the drift gap works as a {\it tiny} time projection chamber and the position of each primary ionization $z_i$ is obtained by knowing the drift velocity of the electrons and the signal time on the strip. The $xz$ points are then fitted with a line $z = ax + b$ and the position\footnote{in the middle of the drift gap, in order to minimize the error on it \cite{mu-TPC}.} is then calculated by:
\begin{equation} \label{eq:mu-TPC}
 x = \frac{\frac{gap}{2} - b}{a}
\end{equation}
In the following, the fields of applications and the performances of the two modes are shown.
\subsection{Orthogonal tracks and B $= 0$}
\begin{figure}[htbp]
\centering 
\includegraphics[width=.5\textwidth]{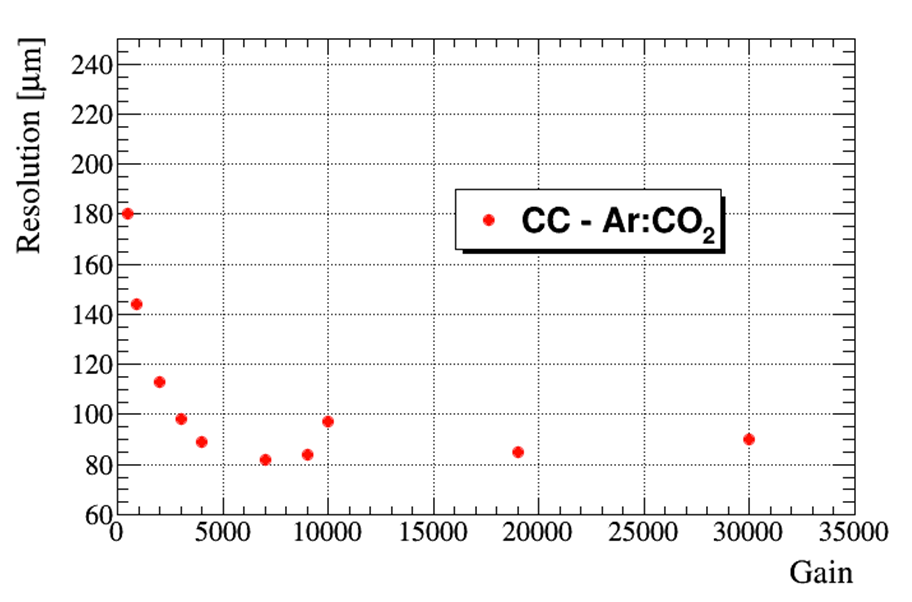} 
\caption{Charge centroid resolution for orthogonal tracks without magnetic field.} \label{fig:res_cc_angle0_B0}
\end{figure}

Events with orthogonal tracks and no magnetic field are the simplest, due to the perfectly Gaussian shape of the charge distribution. Only the diffusion process and the avalanche multiplication apply. The charge centroid is very accurate and resolutions $< 100$$\mu$m have been achieved (see fig.~\ref{fig:res_cc_angle0_B0}). More results can be found in \cite{NIM16}.
\subsection{Inclined tracks and B $\neq 0$}
\begin{figure}[htbp]
\centering 
\includegraphics[width=.45\textwidth]{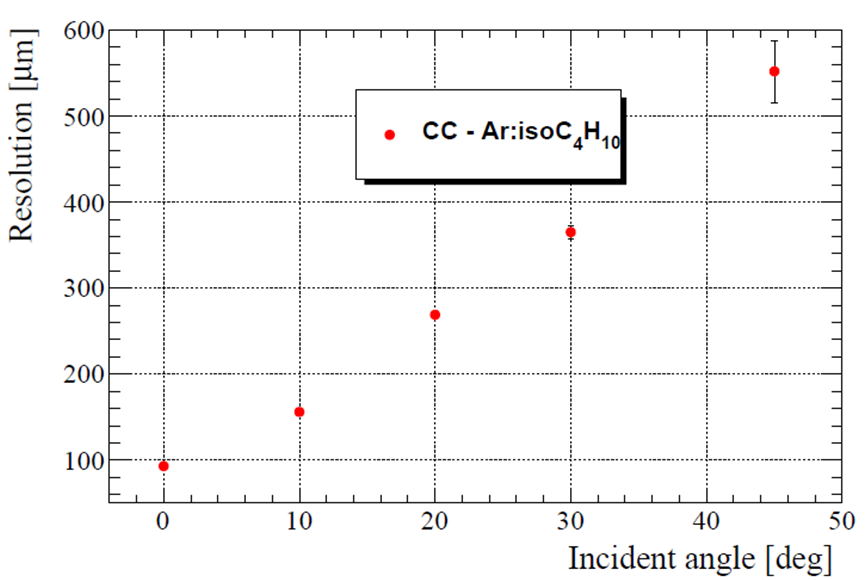} 
\qquad
\includegraphics[width=.45\textwidth]{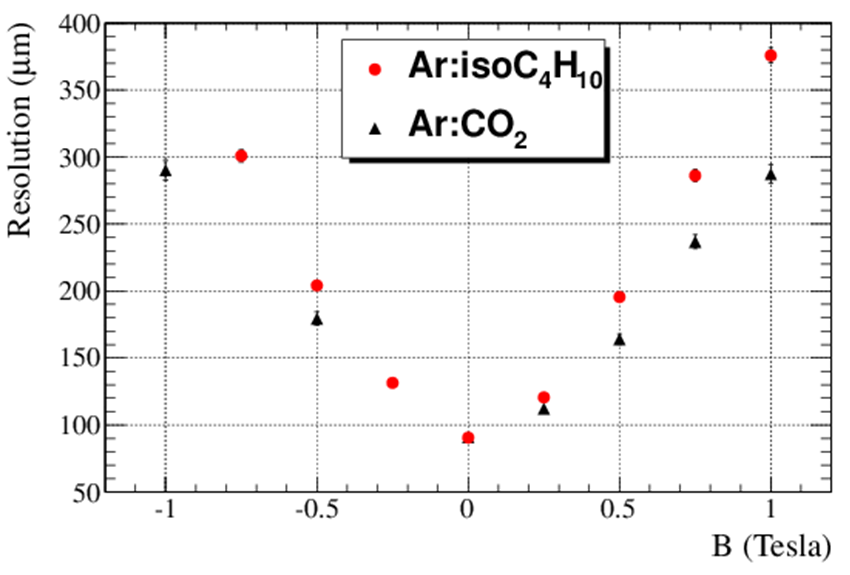}
\caption{Charge centroid resolution worsens for inclined tracks ({\it left}) and in a magnetic field ({\it right}).} \label{fig:res_cc_angle_B}
\end{figure}

\begin{figure}[htbp]
\centering 
\includegraphics[width=.7\textwidth]{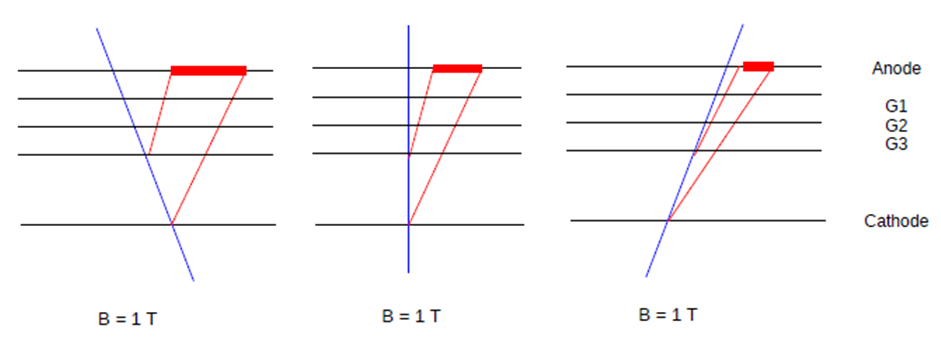} 
\caption{Inclined tracks inside a magnetic field produce a larger/smaller charge distribution.} \label{fig:focus_defocus}
\end{figure}

In a more realistic case, the tracks are inclined with respect to the GEM planes and the whole setup is immersed in a magnetic field. The electrons experience the Lorentz force: this affects the charge distribution shape and the number of firing strips becomes higher. The resolution worsens almost linearly with the incident angle and with the magnetic field (see fig.~\ref{fig:res_cc_angle_B}). 
The situation is critical when both angle $\neq 0$ and B $\neq 0$ at the same time. A {\it focusing} or {\it defocusing} effect appears, depending on the fact that the incident and the Lorentz angles are concordant or discordant (see fig.~\ref{fig:focus_defocus}).
\subsection{Results}
\begin{figure}[htbp]
\centering 
\includegraphics[width=.45\textwidth]{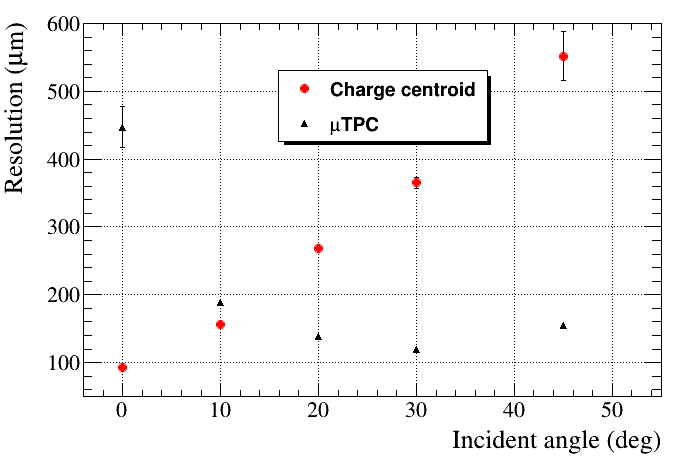} 
\qquad
\includegraphics[width=.45\textwidth]{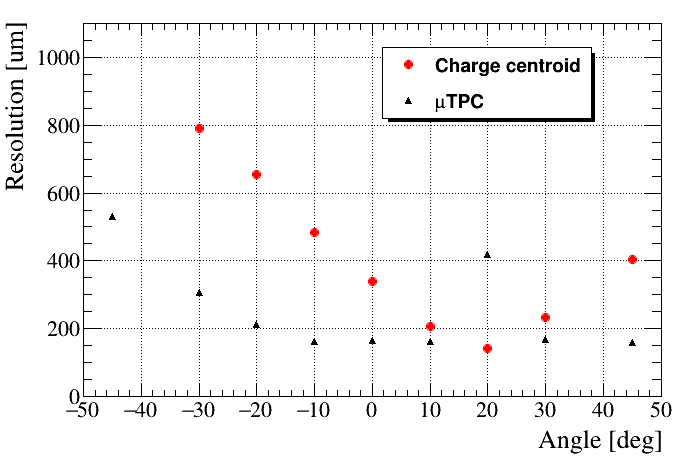}
\caption{Comparison between charge centroid (red dots) and $\mu$-TPC (black dots) in the case of angle $\neq 0$ and B $= 0$ ({\it left}) or B $\neq 0$ ({\it right}). The graphs were obtained in Ar/CO$_2$ where the Lorentz angle is $20^\circ$ \cite{ricIEEE}.} \label{fig:res_cc_mutpc}
\end{figure}

Fig.~\ref{fig:res_cc_mutpc} points out that the charge centroid and the $\mu$-TPC methods are complementary and that their combination can provide the required spatial resolution of $\sim$$130$$\mu$m in $xy$ plane. \\
Layer 2 prototype was tested under extreme conditions, i.e. HV $= 400$V for each GEM foil (corresponding to gain of $10^5$ \cite{triple}) and under a high intensity pion beam, up to some tens of kHz/cm$^2$. It showed great stability and no current peaking problem.
\section{Conclusions}
With the {\bf planar} chambers, efficiencies and resolutions have been evaluated under several working settings (HV, fields). The planar chambers are very stable and show a very good resolution at different incident angles and with magnetic field, provided the use of the charge centroid and $\mu$-TPC alternatively. A composition of the two methods is ongoing to give the best resolution everywhere. \\
Likewise, the {\bf cylindrical} chamber showed high stability under various settings and also under high intensity beam. Tests with a radioactive source and cosmic rays are ongoing to get efficiency and resolution plots comparable to the planar chamber ones. \\
The complete CGEM-IT is under construction and will be completed in $2018$.

\acknowledgments
The research leading to these results has been performed within the BESIIICGEM
Project, funded by the European Commission in the call H2020-MSCA-RISE-2014.


\end{document}